\documentclass[]{article}

\usepackage{amssymb}
\usepackage{amsmath}
\allowdisplaybreaks

\begin{document}

\title{1+1+2 covariant formulation of light propagation in spacetime}
\author{Krzysztof G\l\'od\\ \small Astronomical Observatory, Jagiellonian University,\\ \small Orla 171, 30-244 Krak\'ow, Poland}
\date{}
\maketitle
\begin{abstract}
We present a covariant approach to the problem of light beam propagation in a spacetime. We develop our considerations within the framework of classical geometric optics in general relativity. Using the concept of a~screen surface orthogonal to the observer velocity and to the bundle of geodesics, we introduce covariant four-dimensional definitions for Sachs and Jacobi optical fields and for the area distance. Then we give relationships between them and derive their propagation equations together with initial conditions for these equations. Ultimately, for practical use, we transform the resulting formulas into the redshift-dependent form.
\end{abstract}

\section{Introduction}

The essentials of light propagation in spacetime were developed in Refs. \cite{2013GReGr..45.2691J,1961RSPSA.264..309S}, where the geometry of congruence of null geodesics was considered. In this approach, with the usage of a pseudoorthonormal tetrad which is parallelly transported along the rays, one defines the optical scalars which characterize the rate of change of the geometry of a cross section of the bundle. The propagation equations for the optical scalars are subsequently derived from the Ricci identity. The importance of the optical scalars comes from the fact that the optical expansion rate is directly connected with the area distance. Another approach presented in Refs. \cite{1973lsss.book.....H,1991MNRAS.251..600B,1992grle.book.....S,1993PThPh..90..753S,1994CQGra..11.2345S} is based on the linearity of the geodesic deviation equation which is obeyed by the connecting vectors, which relate neighboring rays in the bundle. Solutions to this equation are coupled to initial conditions through the matrix whose determinant yields the area distance. Both approaches are theoretically equivalent, but computationally they provide two distinct ways for obtaining the area distance. One may find a recent treatment and comprehensive review of the topic in Refs. \cite{2004LRR.....7....9P,2011JCAP...07..008G,2012MNRAS.426.1121C,2013CQGra..30f5020R,2013JCAP...11..019F,2015arXiv151103702F,2016JCAP...09..046Y,2018JCAP...02..015H,2018JCAP...03..012K,2018JCAP...06..040I,2019PhRvD..99f4038G}.

The quantities characterizing the given light beam which are measured by the observer lie in the projective screen space of its 4-velocity and the beam's spatial direction vector. This space is effectively two dimensional; thus, in the standard tetrad-based description of the light beam propagation, one introduces appropriate quantities in the form of complex two-scalars and complex two-dimensional matrices, which makes the equations considerably simpler \cite{1966gref.inco.....P}. In order to solve these equations in the case of some particular spacetime, one nevertheless needs to explicitly construct the tetrad vectors along the considered beam.

Here we provide an alternative description which is devised entirely with the coordinate system of a given spacetime. This formulation of the light propagation fully utilizes the notion of the observer's screen space. It enables the spacetime fields to be covariantly split into parts which are parallel to the observer's 4-velocity or to the beam's spatial direction, or otherwise orthogonal to both of these vectors. This method is similar to the temporal-spatial splitting known from cosmology \cite{1993GReGr..25.1225E,2009GReGr..41..581E}.

\section{Formulation}

\subsection{Screen surface}

We consider some spacetime given by a metric field $g{_m}{_n}$ and some observer with a 4-velocity vector $u{_n}$. The 4-velocity is normalized as
\begin{equation}
u{^a}u{_a}=-1.
\end{equation}
The projection field $P{_m}{_n}$ which projects perpendicularly onto the space orthogonal to the 4-velocity is then defined as
\begin{equation}
P{_m}{_n}=u{_m}u{_n}+g{_m}{_n}.
\end{equation}
Spatial fields are those which are completely orthogonal to the observer's 4-velocity. The effective volume element of the space is represented by the volume field
\begin{equation}
A{_l}{_m}{_n}=u{^a}A{_l}{_m}{_n}{_a},
\end{equation}
where $A{_k}{_l}{_m}{_n}$ is the alternating, totally antisymmetric field of spacetime (we follow the sign conventions used in Ref. \cite{2012reco.book.....E}). The volume field obeys the identity
\begin{equation}
A{^i}{^j}{^k}A{_l}{_m}{_n}=3!P{^{[i}}{_l}P{^j}{_m}P{^{k]}}{_n}.
\end{equation}

We adopt the geometric approximation to the optics in spacetime. The electromagnetic waves perceived by the observer are nearly plane, monochromatic, and short compared with the typical radius of curvature of spacetime. The light propagates along rays whose tangent vector, called the wave vector $k{_n}$, is null, irrotational, and obeys the geodesic equation
\begin{equation}
k{^a}k{_a}=0,\qquad \nabla{_m}k{_n}=\nabla{_n}k{_m},\qquad \dot{k}{_n}=0,
\end{equation}
where the dot denotes $\dot{X}\equiv k{^a}\nabla{_a}X$. This means that the light rays are potential null geodesics.

The circular frequency of the wave $\omega$ measured by the observer is defined as
\begin{equation}
\omega=-u{^a}k{_a}.
\end{equation}
Concurrently, the spatial direction of the wave $d{_n}$ with respect to the observer is given by the unit vector
\begin{equation}
d{_n}=\frac{1}{\omega}P{_n}{^a}k{_a}.
\end{equation}
From these definitions, there follows the decomposition of the wave vector
\begin{equation}
k{_n}=u{_n}\omega+d{_n}\omega.
\end{equation}

The screen field $S{_m}{_n}$ is defined as a symmetric field projecting onto the surface orthogonal to the observer's 4-velocity and to the wave's spatial direction vector. These conditions yield
\begin{equation}
S{_m}{_n}=-d{_m}d{_n}+P{_m}{_n}.
\end{equation}
We shall notice that the screen surface is also orthogonal to the full wave vector. This surface is effectively two dimensional. For a given wave vector, all relevant quantities measurable by the observer are contained in the screen surface. On-screen fields are completely orthogonal simultaneously to the observer's 4-velocity and to the wave's spatial direction vector.

Additionally, we define the area field $A{_m}{_n}$ as a totally antisymmetric field on the screen surface,
\begin{equation}
A{_m}{_n}=d{^a}A{_m}{_n}{_a}.
\end{equation}
It represents the effective area element on the screen surface. It possesses a~property
\begin{equation}
A{^k}{^l}A{_m}{_n}=2!S{^{[k}}{_m}S{^{l]}}{_n},
\end{equation}
which is useful for simplifying formulas.

\subsection{Optical fields}

Let us consider an infinitesimal light beam consisting of close geodesics with a~wave vector $k{_n}$. The change rate of morphology of the beam's screen section is described by the optical deformation rate field $D{_m}{_n}$:
\begin{equation}
D{_m}{_n}=S{_m}{^b}S{_n}{^a}\nabla{_b}k{_a}.
\end{equation}
It is an on-screen gradient of the wave vector. This field is symmetric, since the wave vector is irrotational. It could be further decomposed into its trace-free and pure-trace parts as
\begin{equation}
D{_m}{_n}=\Sigma{_m}{_n}+\frac{1}{2}S{_m}{_n}\Theta,\qquad \Sigma{^a}{_a}=0.
\end{equation}
The traceless field $\Sigma{_m}{_n}$ is the optical shear rate, and it represents the change rate of the shape of the beam's screen section, which could evolve from a circular to an elliptical one. The scalar $\Theta$ is the optical expansion rate, and it represents the change rate of the size of the beam's screen section, which could isotropically expand or contract. These two fields are called the Sachs optical fields. From the above definitions, we have explicitly
\begin{equation}
\Sigma{_m}{_n}=D{_m}{_n}-\frac{1}{2}S{_m}{_n}\Theta,\qquad \Theta=D{^a}{_a}.
\end{equation}
It can be shown (see the Appendix) that the square of the optical shear rate is non-negative:
\begin{equation}\label{squ}
\Sigma{^b}{^a}\Sigma{_b}{_a}\ge0,
\end{equation}
its cube vanishes,
\begin{equation}\label{cub}
\Sigma{^c}{^b}\Sigma{_c}{^a}\Sigma{_b}{_a}=0,
\end{equation}
and moreover, the quantity $\Sigma{_m}{^a}\Sigma{_n}{_a}$ has only the pure-trace part:
\begin{equation}\label{tra}
\Sigma{_m}{^a}\Sigma{_n}{_a}=\frac{1}{2}S{_m}{_n}\Sigma{^b}{^a}\Sigma{_b}{_a}.
\end{equation}
These identities help to reduce the equations containing the optical shear rate.

The transport equations for optical fields along the considered beam are obtained from the Ricci identity for the wave vector,
\begin{equation}
\nabla{_l}\nabla{_m}k{_n}-\nabla{_m}\nabla{_l}k{_n}=R{_l}{_m}{_n}{^a}k{_a},
\end{equation}
where $R{_k}{_l}{_m}{_n}$ is the Riemann tensor. After suitable projections with the usage of the property in Eq. (\ref{tra}), we get two coupled equations:
\begin{align}
S{_m}{^b}S{_n}{^a}\dot{\Sigma}{_b}{_a}=&-\Sigma{_m}{_n}\Theta-S{_m}{^d}k{^c}S{_n}{^b}k{^a}C{_d}{_c}{_b}{_a},\label{sig}\\
\dot{\Theta}=&-\Sigma{^b}{^a}\Sigma{_b}{_a}-\frac{1}{2}\Theta^2-k{^b}k{^a}R{_b}{_a},\label{the}
\end{align}
where $C{_k}{_l}{_m}{_n}$ is the Weyl tensor and $R{_m}{_n}$ is the Ricci tensor. We can see that focusing is directly produced by the trace-free part of the Ricci curvature, but also is indirectly influenced by the conformal curvature through shearing. These equations are called the Sachs optical equations. In practice, we do not solve this system of equations, because we cannot give initial conditions for them at the observation event. Since this point is a vertex where all the beam's rays intersect, the optical expansion rate is singular there.

Usually, the optical equations are presented using the Sachs basis vectors which span the screen surface. Here, we present them using the screen field itself. In the literature, one may find the optical equations derived in this form in, e.g., Refs. \cite{2014CQGra..31t5001U,2015JCAP...10..057L}. The presence of the term with the Weyl tensor in Eq. (\ref{sig}) prevents us from transforming this tensor transport equation to the scalar form for the square of the optical shear rate ($\Sigma{^b}{^a}\Sigma{_b}{_a}$).

\subsection{Area distance}

The actual morphology of the beam's screen section is characterized by the on-screen Jacobi field $J{_m}{_n}$, which is defined by the equation
\begin{gather}
S{_m}{^b}S{_n}{^a}\dot{J}{_b}{_a}=D{_m}{^a}J{_a}{_n},\\
u{^a}J{_n}{_a}=u{^a}J{_a}{_n}=0,\qquad k{^a}J{_n}{_a}=k{^a}J{_a}{_n}=0.
\end{gather}
In general, it is not symmetric. We see that its logarithmic on-screen derivative gives the optical deformation rate field. The Jacobi field encodes the Jacobi matrix of the map relating the physical separations of rays within the beam with the angular separations of these rays seen on the observer's celestial sphere. In particular, the determinant of the Jacobi field $J$ is the Jacobian of the mentioned map, which is the ratio of the physical area of the beam's screen section to its observed solid angle. This enables us to define the area distance $\Delta$ from the observer to the source as the square root of the determinant of the Jacobi field:
\begin{equation}
\Delta^2=J.
\end{equation}
The determinant of the Jacobi field can be calculated with the help of the area field from
\begin{equation}
A{_m}{_n}J=A{^b}{^a}J{_m}{_b}J{_n}{_a},
\end{equation}
which gives
\begin{equation}
J=\frac{1}{2}\bigl(J{^b}{_b}J{^a}{_a}-J{^a}{^b}J{_b}{_a}\bigr).
\end{equation}
This result establishes the covariant formula for the area distance.

We cannot use the definition for direct calculation of the Jacobi field, because the optical deformation rate field is singular at the vertex. Instead, the propagation equation for the Jacobi field is obtained from its definition by differentiation:
\begin{equation}
S{_m}{^b}S{_n}{^a}\ddot{J}{_b}{_a}=-S{_m}{^d}k{^c}S{^e}{^b}k{^a}R{_d}{_c}{_b}{_a}J{_e}{_n}.
\end{equation}
This equation is a reminiscence of the geodesic deviation equation which holds for the Jacobi vectors connecting nearby rays in the beam. The factor in the term on the right-hand side of the equation is called the optical tidal field. Once we solve this equation, we can find the area distance.

Alternatively, we can return to the Sachs optical equations. The optical expansion rate is expressed by the area distance as follows:
\begin{equation}
\Theta=2\frac{\dot{\Delta}}{\Delta}.
\end{equation}
Hence, we rewrite the Sachs optical equations into the form
\begin{align}
S{_m}{^b}S{_n}{^a}\dot{\Xi}{_b}{_a}=&-\Delta^2S{_m}{^d}k{^c}S{_n}{^b}k{^a}C{_d}{_c}{_b}{_a},\label{she}\\
\ddot{\Delta}=&-\frac{1}{2}\frac{1}{\Delta^3}\Xi{^b}{^a}\Xi{_b}{_a}-\frac{1}{2}\Delta k{^b}k{^a}R{_b}{_a},\label{dis}
\end{align}
where we have introduced the scaled optical shear rate $\Xi{_m}{_n}=\Delta^2\Sigma{_m}{_n}$. This system of equations can be solved to obtain the area distance directly. Equation (\ref{she}) shows that the necessary source for the optical shearing is a nonzero conformal curvature. Since $\Xi{^b}{^a}\Xi{_b}{_a}\ge0$, while neglecting the optical shearing in Eq. (\ref{dis}), we overestimate the distance.

\subsection{Initial conditions}

In order to impose the initial conditions for the considered equations, one needs to know the relation between the scaled optical shear rate and the Jacobi field:
\begin{equation}
\Xi{_m}{_n}=-\Delta S{_m}{^b}S{_n}{^a}\bigl(J{_b}{_a}\Bigl(\frac{J{^c}{_c}}{\Delta}\dot{\Bigr)}-J{_b}{^c}\Bigl(\frac{J{_c}{_a}}{\Delta}\dot{\Bigr)}\bigr).
\end{equation}
Since the observation event is a vertex point for the beam's rays, the Jacobi field vanishes there:
\begin{equation}
J{_m}{_n}\big|_0=0.
\end{equation}
By the relationships between the respective fields, this implies sequentially that
\begin{gather}
\Delta\big|_0=0,\qquad \frac{\Xi{_m}{_n}}{\Delta^2}\big|_0=0,\\
\Xi{_m}{_n}\big|_0=0,\qquad \frac{J{_m}{_n}}{\Delta}\big|_0=\frac{\dot{J}{_m}{_n}}{\dot{\Delta}}\big|_0,
\end{gather}
and additionally there follows the identity between initial conditions for derivatives of the area distance and the Jacobi field:
\begin{equation}
\dot{\Delta}^2\big|_0=\frac{1}{2}\bigl(\dot{J}{^b}{_b}\dot{J}{^a}{_a}-\dot{J}{^a}{^b}\dot{J}{_b}{_a}\bigr)\big|_0.
\end{equation}
We can see that the scaled optical shear rate goes to zero faster than the area distance. Because of its properties and symmetries, we shall impose the initial conditions for only two of the components of the scaled optical shear rate. Likewise, we give the initial conditions for four of the components of the Jacobi field.

The initial condition for the derivative of the area distance comes from the physical requirement that in the vicinity of the vertex, the distance should correspond to the path traveled by the photon with respect to the observer. If $S$ is the affine parameter along the geodesic $x{^n}$ crossing the vertex, then the infinitesimal distance $dl$ from the observer in the direction of the source can be estimated as
\begin{equation}
dl=-d{_a}dx{^a}=-d{_a}k{^a}dS=-\omega dS.
\end{equation}
Hence, this gives
\begin{equation}
\dot{\Delta}\big|_0=-\omega\big|_0\equiv-\omega_0.
\end{equation}
The minus sign comes due to the choice that the wave vector is future-oriented. The initial conditions for the components of the derivative of the Jacobi field are subjected only to the identity mentioned above, and otherwise, they are unrestricted.

\subsection{Redshift dependence}

The area distance is determined by the derived equations as a function of the affine parameter along the given geodesic. Since the affine parameter is not observable, it is useful to introduce the redshift $Z$ as a new independent variable
\begin{equation}
Z=\frac{\omega}{\omega_0}-1.
\end{equation}
Its differential connection with the affine parameter reads
\begin{equation}
k{^a}\partial{_a}=-\omega_0\frac{1}{\mathfrak{B}}l{^a}\partial{_a},\qquad \mathfrak{B}=l{^b}l{^a}\nabla{_b}u{_a},
\end{equation}
where we have introduced the new null vector $l{^n}=\frac{dx{^n}}{dZ}$. This relation could be obtained by the calculation of the derivative of the redshift with respect to the affine parameter. Accordingly, the second derivative reads
\begin{gather}
k{^b}\partial{_b}\bigl(k{^a}\partial{_a}\bigr)=\omega_0^2\frac{1}{\mathfrak{B}^2}l{^b}\partial{_b}\bigl(l{^a}\partial{_a}\bigr)+\omega_0^2\frac{\mathfrak{C}}{\mathfrak{B}^3}l{^a}\partial{_a},\\
\mathfrak{C}=l{^c}l{^b}l{^a}\nabla{_c}\nabla{_b}u{_a}.
\end{gather}
Therefore, the geodesic equation in the redshift-dependent form can be written as
\begin{equation}
l'{_n}=-\frac{\mathfrak{C}}{\mathfrak{B}}l{_n},
\end{equation}
where the prime denotes $X'\equiv l{^a}\nabla{_a}X$. This equation enables us to find the geodesic curve directly as a function of the redshift.

In the redshift-dependent approach, the circular frequency is absent from the equations. However, it can be recovered as
\begin{equation}
\omega=\omega_0\frac{\mathfrak{A}}{\mathfrak{B}},\qquad \mathfrak{A}=l{^a}u{_a}.
\end{equation}
To be consistent with the derivation, while specifying the initial conditions for the above geodesic equation, one should assure that
\begin{equation}
\mathfrak{B}\big|_0=\mathfrak{A}\big|_0,
\end{equation}
which sets the initial normalization for the components of the vector $l{_n}$.

The system of equations for the area distance in the redshift-dependent form reads
\begin{align}
S{_m}{^b}S{_n}{^a}X'{_b}{_a}=&-\frac{1}{\mathfrak{B}}\Delta^2S{_m}{^d}l{^c}S{_n}{^b}l{^a}C{_d}{_c}{_b}{_a},\\
\Delta''=&-\frac{\mathfrak{C}}{\mathfrak{B}}\Delta'-\frac{\mathfrak{B}^2}{2}\frac{1}{\Delta^3}X{^b}{^a}X{_b}{_a}-\frac{1}{2}\Delta l{^b}l{^a}R{_b}{_a},
\end{align}
where we have rescaled the optical shear rate as $X{_m}{_n}=-\frac{1}{\omega_0}\Xi{_m}{_n}$. The initial condition for the derivative of the area distance with respect to the redshift takes the form
\begin{equation}
\Delta'\big|_0=\mathfrak{B}\big|_0.
\end{equation}
Finally, the equation for the Jacobi field as a function of the redshift reads
\begin{equation}
S{_m}{^b}S{_n}{^a}J''{_b}{_a}=-\frac{\mathfrak{C}}{\mathfrak{B}}S{_m}{^b}S{_n}{^a}J'{_b}{_a}-S{_m}{^d}l{^c}S{^e}{^b}l{^a}R{_d}{_c}{_b}{_a}J{_e}{_n},
\end{equation}
where we have used the fact that
\begin{equation}
\mathfrak{B}'=-\mathfrak{C}.
\end{equation}
The relation between initial derivatives holds in the form
\begin{equation}
\Delta'^2\big|_0=\frac{1}{2}\bigl(J'{^b}{_b}J'{^a}{_a}-J'{^a}{^b}J'{_b}{_a}\bigr)\big|_0.
\end{equation}
These equations could be easily implemented and numerically solved to get solutions as functions of the redshift unmediated by the affine parameter.

\section{Summary}

We have presented the problem of light propagation in a narrow beam in terms of the 1+1+2 splitting of spacetime. In this approach, the spacetime along the beam is covariantly split by the observer's 4-velocity vector and the beam's spatial direction vector into its temporal, radial, and screen parts. This enables us to give consistent and covariant definitions for basic quantities characterizing properties of the propagating light beam, particularly for the area distance. The practical advantage of this splitting-based approach over the standard tetrad-based approach is that one may avoid the construction of the Sachs basis vectors along the beam and proceed entirely within the full four-dimensional formalism. Moreover, when dealing with some theoretical concepts concerning the light propagation in spacetime, like for example weak lensing, redshift drift, or gauge-invariant perturbation theory, it is more appealing and elegant to work with covariant fields than matrices.

Within the developed formulation, we have recalled and elaborated upon two analytically equivalent but numerically independent methods for determination of the distance function, one from the Sachs optical fields and another from the Jacobi field. In a situation of particular spacetime, they both can be used to compare their outcomes or efficiency.

The formulation presented here is complementary to the existing two-dimen\-sional approaches. As is seen, however, it is mainly oriented on the observer's measurements rather than on the beam's intrinsic properties. Because of this, it could be especially useful in applications to cosmology for studies of light beam propagation in possibly inhomogeneous cosmological models. In regard to this, we have covariantly translated the developed formulation into the redshift-dependent form, which for example makes it possible to determine the area distance directly as a function of the redshift.

\appendix

\section{Appendix}

Since on-screen fields are effectively two dimensional, rank-2 on-screen fields can be expressed as a product of two on-screen vector fields. These vector fields can be in turn decomposed in a basis of two mutually orthonormal on-screen vector fields---let us denote them $A{_n}$ and $B{_n}$---for which we have
\begin{gather}
u{^a}A{_a}=d{^a}A{_a}=0,\qquad u{^a}B{_a}=d{^a}B{_a}=0,\\
A{^a}B{_a}=0,\qquad A{^a}A{_a}=B{^a}B{_a}=1.
\end{gather}
For the screen field $S{_m}{_n}$, which satisfies $S{_m}{_n}=S{_n}{_m}$, $S{^a}{_a}=2$, $S{_m}{^a}S{_n}{_a}=S{_m}{_n}$, this implies that
\begin{equation}
S{_m}{_n}=A{_m}A{_n}+B{_m}B{_n}.
\end{equation}
On the other hand, for the optical shear rate $\Sigma{_m}{_n}$, which is symmetric, $\Sigma{_m}{_n}=\Sigma{_n}{_m}$, and traceless, $\Sigma{^a}{_a}=0$, this decomposition gives in general that
\begin{equation}
\Sigma{_m}{_n}=\alpha\bigl(A{_m}A{_n}-B{_m}B{_n}\bigr)+\beta\bigl(A{_m}B{_n}+B{_m}A{_n}\bigr),
\end{equation}
where $\alpha$ and $\beta$ are some scalar fields. Using the above representation, it is straightforward to verify that Eqs. (\ref{squ}), (\ref{cub}), and (\ref{tra}) are fulfilled.

\bibliographystyle{unsrt}
\bibliography{art}

\begin{thebibliography}{10}

\bibitem{2013GReGr..45.2691J}
P.~{Jordan}, J.~{Ehlers}, and R.~K. {Sachs}.
\newblock {Republication of: Contributions to the theory of pure gravitational
  radiation. Exact solutions of the field equations of the general theory of
  relativity II}.
\newblock {\em General Relativity and Gravitation}, 45:2691--2753, December
  2013.

\bibitem{1961RSPSA.264..309S}
R.~{Sachs}.
\newblock {Gravitational Waves in General Relativity. VI. The Outgoing
  Radiation Condition}.
\newblock {\em Proceedings of the Royal Society of London Series A},
  264:309--338, November 1961.

\bibitem{1973lsss.book.....H}
S.~W. {Hawking} and G.~F.~R. {Ellis}.
\newblock {\em {The large-scale structure of space-time.}}
\newblock Cambridge University Press, Cambridge, 1973.

\bibitem{1991MNRAS.251..600B}
R.~D. {Blandford}, A.~B. {Saust}, T.~G. {Brainerd}, and J.~V. {Villumsen}.
\newblock {The distortion of distant galaxy images by large-scale structure}.
\newblock {\em Monthly Notices of the Royal Astronomical Society},
  251:600--627, August 1991.

\bibitem{1992grle.book.....S}
P.~{Schneider}, J.~{Ehlers}, and E.~E. {Falco}.
\newblock {\em {Gravitational Lenses}}.
\newblock Springer-Verlag, Berlin, 1992.

\bibitem{1993PThPh..90..753S}
M.~{Sasaki}.
\newblock {Cosmological Gravitational Lens Equation --- Its Validity and
  Limitation ---}.
\newblock {\em Progress of Theoretical Physics}, 90:753--781, October 1993.

\bibitem{1994CQGra..11.2345S}
S.~{Seitz}, P.~{Schneider}, and J.~{Ehlers}.
\newblock {Light propagation in arbitrary spacetimes and the gravitational lens
  approximation}.
\newblock {\em Classical and Quantum Gravity}, 11:2345--2373, September 1994.

\bibitem{2004LRR.....7....9P}
V.~{Perlick}.
\newblock {Gravitational Lensing from a Spacetime Perspective}.
\newblock {\em Living Reviews in Relativity}, 7:9, September 2004.

\bibitem{2011JCAP...07..008G}
M.~{Gasperini}, G.~{Marozzi}, F.~{Nugier}, and G.~{Veneziano}.
\newblock {Light-cone averaging in cosmology: formalism and applications}.
\newblock {\em Journal of Cosmology and Astroparticle Physics}, 7(2011):008,
  July 2011.

\bibitem{2012MNRAS.426.1121C}
C.~{Clarkson}, G.~F.~R. {Ellis}, A.~{Faltenbacher}, R.~{Maartens}, O.~{Umeh},
  and J.-P. {Uzan}.
\newblock {(Mis)interpreting supernovae observations in a lumpy universe}.
\newblock {\em Monthly Notices of the Royal Astronomical Society},
  426:1121--1136, October 2012.

\bibitem{2013CQGra..30f5020R}
P.~H.~F. {Reimberg} and L.~R. {Abramo}.
\newblock {The Jacobi map for gravitational lensing: the role of the
  exponential map}.
\newblock {\em Classical and Quantum Gravity}, 30(6):065020, March 2013.

\bibitem{2013JCAP...11..019F}
G.~{Fanizza}, M.~{Gasperini}, G.~{Marozzi}, and G.~{Veneziano}.
\newblock {An exact Jacobi map in the geodesic light-cone gauge}.
\newblock {\em Journal of Cosmology and Astroparticle Physics}, 11(2013):019,
  November 2013.

\bibitem{2015arXiv151103702F}
P.~{Fleury}.
\newblock {Light propagation in inhomogeneous and anisotropic cosmologies}.
\newblock {\em arXiv e-prints}, pages gr--qc/1511.03702, November 2015.

\bibitem{2016JCAP...09..046Y}
J.~{Yoo} and F.~{Scaccabarozzi}.
\newblock {Unified treatment of the luminosity distance in cosmology}.
\newblock {\em Journal of Cosmology and Astroparticle Physics}, 9(2016):046,
  September 2016.

\bibitem{2018JCAP...02..015H}
C.~{Hellaby} and A.~{Walters}.
\newblock {Calculating observables in inhomogeneous cosmologies. Part I:
  general framework}.
\newblock {\em Journal of Cosmology and Astroparticle Physics}, 2(2018):015,
  February 2018.

\bibitem{2018JCAP...03..012K}
M.~{Korzy{\'n}ski} and J.~{Kopi{\'n}ski}.
\newblock {Optical drift effects in general relativity}.
\newblock {\em Journal of Cosmology and Astroparticle Physics}, 3(2018):012,
  March 2018.

\bibitem{2018JCAP...06..040I}
D.~{Ivanov}, S.~{Liberati}, M.~{Viel}, and M.~{Visser}.
\newblock {Non-perturbative results for the luminosity and area distances}.
\newblock {\em Journal of Cosmology and Astroparticle Physics}, 6(2018):040,
  June 2018.

\bibitem{2019PhRvD..99f4038G}
M.~{Grasso}, M.~{Korzy{\'n}ski}, and J.~{Serbenta}.
\newblock {Geometric optics in general relativity using bilocal operators}.
\newblock {\em Physical Review D}, 99(6):064038, March 2019.

\bibitem{1966gref.inco.....P}
R.~{Penrose}.
\newblock {General-Relativistic Energy Flux and Elementary Optics}.
\newblock In B.~{Hoffmann}, editor, {\em Perspectives in Geometry and
  Relativity}. Indiana University Press, Bloomington, 1966.

\bibitem{1993GReGr..25.1225E}
J.~{Ehlers}.
\newblock {Contributions to the relativistic mechanics of continuous media}.
\newblock {\em General Relativity and Gravitation}, 25:1225--1266, December
  1993.

\bibitem{2009GReGr..41..581E}
G.~F.~R. {Ellis}.
\newblock {Republication of: Relativistic cosmology}.
\newblock {\em General Relativity and Gravitation}, 41:581--660, March 2009.

\bibitem{2012reco.book.....E}
George F.~R. {Ellis}, Roy {Maartens}, and Malcolm A.~H. {MacCallum}.
\newblock {\em {Relativistic Cosmology}}.
\newblock Cambridge University Press, Cambridge, 2012.

\bibitem{2014CQGra..31t5001U}
O.~{Umeh}, C.~{Clarkson}, and R.~{Maartens}.
\newblock {Nonlinear relativistic corrections to cosmological distances,
  redshift and gravitational lensing magnification: II. Derivation}.
\newblock {\em Classical and Quantum Gravity}, 31(20):205001, October 2014.

\bibitem{2015JCAP...10..057L}
M.~{Lavinto} and S.~{R{\"a}s{\"a}nen}.
\newblock {CMB seen through random Swiss Cheese}.
\newblock {\em Journal of Cosmology and Astroparticle Physics}, 10(2015):057,
  October 2015.

\end{thebibliography}

\end{document}